# A Lightweight Pipeline for Noisy Speech Voice Cloning and Accurate Lip-Sync Synthesis

**Javeria Amir[1], Farwa Attaria[1], Mah Jabeen[1], Umara Noor[1], Zahid Rashid[2]**

[1]Department of Software Engineering, Faculty of Computing and Information Technology, International Islamic University, Islamabad, Pakistan

[2]Technology Management Economics and Policy Program, College of Engineering, Seoul National University, 1 Gwanak-Ro, Gwanak-Gu, 08826 Seoul, South Korea

farwa.bsse4211@iiu.edu.pk, jaweria.bsse4195@iiu.edu.pk, mah.bsse4191@iiu.edu.pk, umara.zahid@iiu.edu.pk, rashidzahid@snu.ac.kr

Corresponding Author: Umara Noor

## Abstract

Recent developments in voice cloning and talking-head generation demonstrate impressive capabilities in synthesizing natural speech and realistic lip synchronization. Current methods typically require and are trained on large-scale datasets and computationally intensive processes using clean, studio-recorded inputs, which is infeasible in noisy or low-resource environments. In this paper, we introduce a new modular pipeline comprising Tortoise text to speech, a transformer-based latent diffusion model that can perform high-fidelity zero-shot voice cloning given only a few training samples, and Wav2Lip, a lightweight generative adversarial network architecture for robust real-time lip synchronization. The solution will contribute to many essential tasks concerning less reliance on massive pretraining, generation of emotionally expressive speech, and lip-sync in noisy and unconstrained scenarios. In addition, the modular structure of the pipeline allows an easy extension for future multimodal and text-guided voice modulation, and it could be used in real-world systems. Our experimental results show that the proposed system produces competition-level sound quality and lip-sync with a much smaller computational cost, indicating the possibility of deploying it in resource-constrained scenarios.

## Keywords

# 1. Introduction

Voice clone and talking head generation systems have made tremendous progress in the past few years, benefiting from the development of deep and generative models. These devices can be employed for virtual assistants, entertainment, telepresence, and assistive communication, making human-computer interaction more realistic and personalized, based on interactive and audio-visual context. Despite advancements, the state-of-the-art solutions heavily rely on big data and sophisticated computational resources and therefore may not be practical for real-world low-resource or noisy settings. Furthermore, a number of existing models have difficulty achieving expressive and emotionally rich voice synthesis along with accurate and efficient lip-syncing, particularly in real time.

The focus of this work is the challenge of creating an effective and practical pipeline that can perform zero-shot voice cloning and real-time lip sync from a small noise-corrupted input. While previous works including OpenVoice [1], MiniMax-Speech [2], and TalkCLIP [3] have advanced the frontier of speaker fidelity and style controllable synthesis, they generally leverage extensive pretraining, complex pipelines, or specialized hardware. Likewise, talking-head models, such as Diffused Heads [4] or Audio-Driven Talking Face [5], are impressive in quality but not as well-suited to real-time deployment and handling noisy inputs.

Even in recent studies, weak-resource settings, simple yet vivid emotion, and steady lip-sync across noisy audio still get short shrift. To address this, we stack Tortoise (Text To Speech) TTS [6], a few-shot voice-cloner that rides on a transformer diffusion backbone with Wav2Lip [7], a quick Generative Adversarial Network (GAN) [8] that masks and matches mouths in near-real time. This plug-and-play pipeline trims data and computes costs while boosting vocal warmth and picture steadiness, offering an up-to-date tool kit for messy, everyday media projects.

The major study objective is to develop, test, and evaluate a scaled-down video pipeline so that it can be demonstrated as functioning well on inexpensive hardware, retaining voices naturally, and lip reading matching to sound. Experiments confirm that such an approach could influence virtual agents, remote language tutors, as well as helpful avatars for disabled people.

The paper unfolds as follows: Section 2 reviews existing literature and related work; Section 3 describes the proposed method and pipeline design; Section 4 outlines the experimental set-up and results; Section 5 discusses findings and limitations; and Section 6 winds up and delineates future work.

## 2. Literature Review

Advanced voice cloning and talking-head synthesis attained a breakthrough mainly due to deep learning and transformers, as well as generative adversarial networks GAN[8]. In this section, we review the literature based on the following: voice cloning techniques, lip tracking methods, and multi-modal synthesis systems.

### 2.1. Voice Cloning Techniques

Early research in speaker cloning utilized speaker adaptation in TTS systems and needed a large amount of training data from each speaker. Recent methods can learn to clone voices from only a few or zero examples. Jia et al. [9] presented a neural speaker embedding model that supported low-resource speaker adaptation. However, their method suffered in terms of expressive speech synthesis. By utilizing transformer-based latent diffusion, the Tortoise TTS model [6] improves upon this by generating highly natural and expressive voices from a few seconds of reference audio, overcoming prior limitations in emotional richness and style transfer. However, these models still require a lot of computational resources and are hard to use in real-time.

### 2.2. Lip-Syncing Techniques

Lip syncing techniques vary from traditional computer vision techniques to deep learning-based techniques. The Wav2Lip[7] model employed a GAN to generate synchronized lip movement based on an audio signal and a still image. Wav2Lip set a new record for lip-sync accuracy while maintaining a good level of accuracy with noise in the audio signal. Importantly, Wav2Lip was different from other methods in that previous phoneme-aligned networks required processing of the phonemes whereas Wav2Lip worked directly with raw audio. This new methodology also made it more robust to the different quality of sound inherent in the audio signal. Other alternatives, like SyncNet [10], specifically focused on audio-visual correspondence but did not even perform on par with Wav2Lip in real-time. It is important to highlight that even while Wav2Lip can do real-time lip-syncing, no work has been conducted to explore a real-time lip-sync with an expressive TTS model.

### 2.3. Integrated Audio-Visual Generation Systems

Very few works have integrated voice cloning and lip-syncing in a single pipeline. Notable exceptions include TalkCLIP [3] and Diffused Heads [4] which produce talking faces conditioned on audio but rely on large datasets or complex conditioning signals to create new talking face videos, which presents challenges regarding accessibility. MiniMax-Speech [2] provides a multi-style voice cloning process but does not address synchronized video generation. Overall, creating efficient pipelines that can be expressive in voice synthesis generation and effective lip-syncing has not received adequate attention.

### 2.4. Critical Analysis

While voice cloning has come a long way in terms of naturalness and expressiveness, as well as lip-syncing models becoming more accurate and capable of real-time performance, integrating all of these technologies into one cohesive, efficient pipeline for zero-shot voice cloning and talking-head generation is still not common. Most existing systems require large amounts of computational power and/or data, and won't tend to be robust in noisy audio or low-resource scenarios, and very few efforts in integrating talking heads with speech synthesis retain the emotional expressiveness provided in their synthesized speech.

### 2.5. Research Gap

This review points out that we lack a practical modular pipeline for transformer-based expressive voice cloning (e.g., Tortoise TTS) paired with GAN-based real-time lip synchronization (e.g., Wav2Lip) that is optimized for low-resource and noisy input conditions. Our work aims to fill this gap by building and evaluating a system that provides a balance of a faithful voice, expressiveness, and lip-sync accuracy that can be run computationally efficiently enough for practical applications.

## 3. Proposed Methodology

This section describes our proposed system architecture, components, and flow that combines zero-shot voice cloning with lip synchronization in real time. We propose to develop a modular, efficient, and robust pipeline that can produce emotionally expressive synthetic speech when limited samples are provided and lip-synchronize it to any video sample in real time. The architecture is optimized for low-resources and noisy environments that the current state-of-the-art approaches tend not to address.

## 3.1. System Overview

The system we are proposing has two main components. (1) Tortoise TTS [6]: This performs expressive voice cloning from reference audio and text input. (2) Wav2Lip [7]: This generates lip-synced talking head videos from synthesized audio and a reference face video/image. The two components are mapped into a modular pipeline, where we also highlight the input and output data, as well as the limited pre-processing is required.

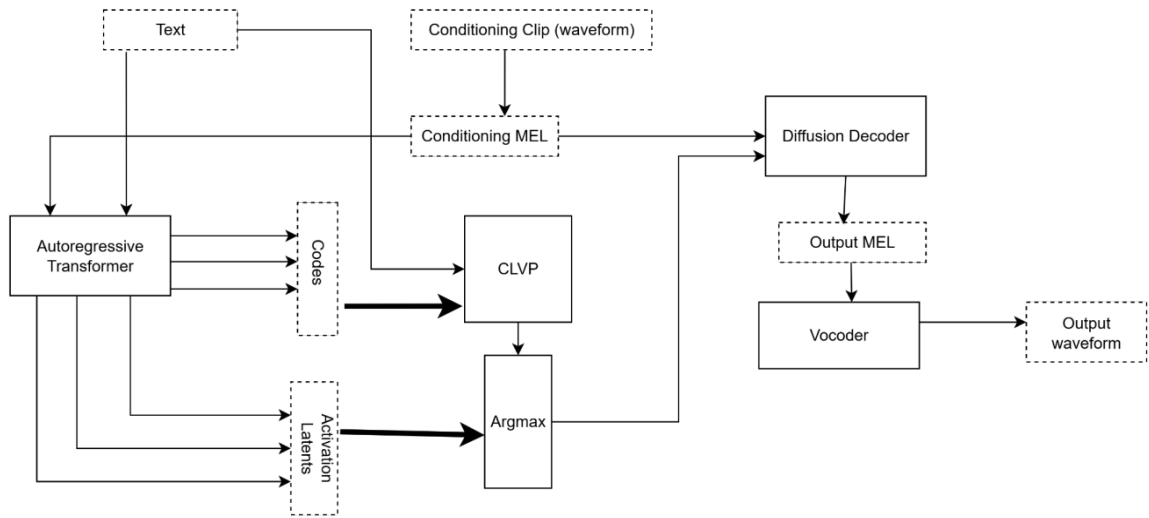

Figure 1 Tortoise TTS architecture

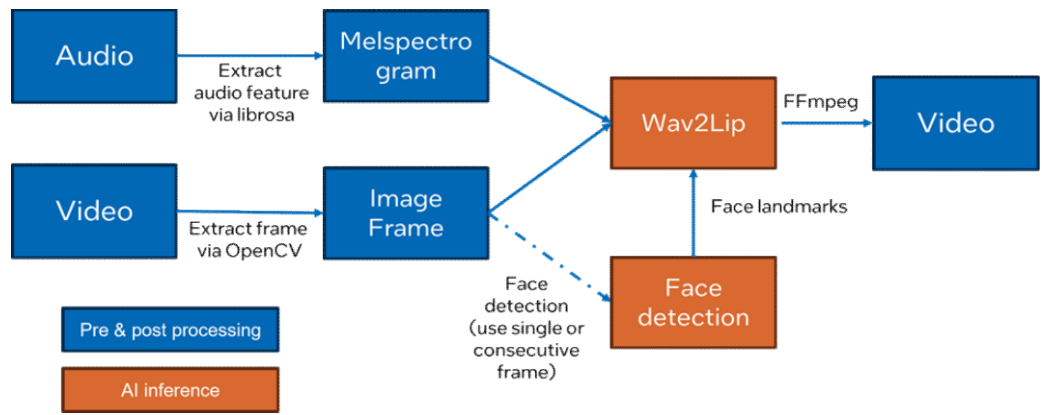

Figure 2 Wav2Lip architecture

As illustrated in Figure 1, the process begins with the short user-supplied voice sample and input text. Tortoise TTS will synthesize high-fidelity speech that matches the style of the reference voice. The resulting audio is then fed into Wav2Lip, along with a still image or video frame of a face, to generate a talking-head video in which the audio and video are well synchronized. This whole system is designed to work quickly enough to operate in real-time or near-real-time conditions and does not require any style tokens, speaker embeddings, or fine-tuning.

### 3.2. Voice Cloning with Tortoise TTS

Tortoise TTS is an auto-regressive transformer-based model utilizing a diffusion decoder. Out of the box, Tortoise TTS supports zero-shot voice cloning that can replicate a target speaker's voice using a short (and sometimes noisy) sample of reference audio. The process works as follows:

    a) **Voice Embedding Extraction:** Takes a reference voice sample (3-10 seconds) and runs through a pre-trained encoder to generate a speaker embedding.

    b) **Text-to-Latent Conversions:** The input text is tokenized, and then goes through an autoregressive transformer modeling standard prosody and timing depending on your choices and what any input sounds like.

    c) **Latent-to-Audio Conversion:** The latent representation is decoded to a waveform audio track using a diffusion-based decoder teaching the model to generate audio that sounds expressive, natural, and human.

The latent diffusion design of Tortoise capitalizes on emotional cues, intonation, and speaker identity to a much greater degree than standard, popular TTS models (like Tacotron or FastSpeech), particularly when inputs are not constrained and/or are noisy.

### 3.3. Real-Time Lip Syncing with Wav2Lip

Wav2Lip[3] is a GAN-based model that provides accurate lip-syncing from raw audio input and video feedback. The Wav2Lip inputs are:

    a)    The synthetic audio produced by Tortoise TTS

    b)    A static or video of a target speaker (face only, no body)

The model uses a synchronization discriminator that was trained on the SyncNet model architecture [10] so that lip movement is temporally aligned to the audio waveform. Wav2Lip runs with low latency and preserves accuracy in challenging conditions such as background noise or expressive speech patterns in the input audio.

Unlike many models for talking-head synthesis in the literature, which require phoneme-level alignment or many speakers, Wav2Lip generalizes to different speakers and facial styles only relying on audio-visual correlation.

### 3.4. Data Flow and Integration

The data flow is designed to be both sequential and modular to allow components to be easily reused or replaced independently. The integration happens in the following way:

    a)    **Input Layer:** The user submits a short voice sample, as well as their target text.

    b)    **Voice Cloning:** Tortoise TTS takes the input text and reference voice to synthesize the expressive speech audio.

    c)    **Lip Syncing:** This creates a steady audio output, which integrates with Wav2Lip, along with a face video to create a synchronized talking-head video.

    d)    **Output:** The output of the video displays a real-world, talking avatar that replicates the voice and lip movement of the user-supplied, voice-sample speaker.

No fine-tuning or domain-specific pre-processing is necessary at runtime, allowing robustness and deployment in multiple environments.

### 3.5. Real-Time and Low-Resource Performance

Certainly, to be useful in real applications, the pipeline has been designed to employ low-latency features. While Tortoise TTS uses GPU acceleration for diffusion decoding, it can also operate with mid-range hardware since it is explicitly modular in its stages. Furthermore, Wav2Lip performs inference in real time and could be deployed to mobile or edge devices without considerable optimization.

Our design priorities are:

- Low sample requirements (voice cloning uses <10 seconds of audio)
- Low pre-processing
- Noise tolerant
- Modular extensibility (e.g., adding emotion control or language switch)

## 4. Dataset and Data Collection

The performance and validity of any AI-based audio-visual synthesis system are dependent on high-quality, diverse, and realistic input data. Although our pipeline was designed to function under low-resource and noisy input conditions, we had data selection strategies that reflected stream usability, not clean studio inputs. This section highlights the datasets used, their qualities, how they were prepared, and the ethical matters we considered.

## 4.1. Data Description

Our system's evaluation was conducted using digital media of Angelina Jolie that had previously been made available or shared with the public as representative speaker data. However, the pipeline for our system can be generalized to any speaker, as long as a short sample of their speech and one clear video clip are provided. We used:

- **Voice Sample:** A single excerpt of speech (10+ seconds) extracted from an interview using a natural and expressive speaking style. The environment included light background noise typical of public media archival recordings.
- **Visual Input:** A short clip was selected of a public talk of the frontal face. The video had no camera shake, consistent lighting, clear lip movements, and sequential expressive speech making it a good option for testing lip synchronization.

The setup uses a real-world use case in which users can have only a limited amount of audio and a short video, or audio content may have been recorded off online streaming services or mobile phones.

## 4.2. Data Collection Process

The voice sample was extracted with FFmpeg [11] and subjected to spectral noise reduction and filtering for clarity and cleaned input to the Tortoise TTS model. The voice sample was not split, trimmed, or edited for silence removal to maintain the natural cadence of speech.

The video was chosen to match the voice sample and to provide visible, expressive facial motion. Only one video was used, and it was preprocessed for Wav2Lip input while retaining lip matching, and visual fidelity in sound.

## 4.3. Data Attributes

The following properties detail the minimum dataset used for testing:

- **Audio:**
    - **Sample:** 1 speech segment
    - **Length:** ~10-15 seconds
    - **Format:** 22.05 kHz mono WAV
    - **Source:** Publicly available interview recording
    - **Preprocessing:** Denoised, normalized
- **Video:**
    - **Sample:** 1 short video

- **Resolution:** 640x360
- **Pose:** Frontal-facing
- **Expression:** Neutral to expressive (clear lip movement)
- **Source:** Publicly available speech footage

These limited inputs were adequate to create expressive talking-head videos that were temporally aligned with the audio, highlighting the efficacy of the pipeline in limited situations.

### 4.4. Preprocessing and Cleaning

The preprocessing required was minimal because an authentic representation of videos was desired to demonstrate the system's tolerability of real-world data points:

- **Audio:** Cleaned and denoised through normal clean-up procedures using spectral filtering; amplitude normalized for the Tortoise TTS to have a stable synthesis.
- **Video:** Cropped and resized within the required parameters presented by Wav2Lip; lip visibility and timing were made to closely resemble audio.

No face landmarking, silence trimming, or augmentation was applied. The goal was to keep the raw expressiveness while still keeping the model's compatibility.

### 4.5. Ethical and Privacy Considerations

All content that was used for evaluation was obtained from publicly accessible sources in compliance with their license and their research/usage policies. No private or personally identifiable information was collected. The system is only intended for research and academic purposes. If this were to be used publicly in the future, explicit consent from the user would need to be collected and ethical safeguards to maintain privacy, data integrity, and responsible use [12].

### 4.6. Data Summary

| Modality | Source | Samples | Attributes Captured |
| --- | --- | --- | --- |
| Voice Sample | Public interview | 1 | Emotional tone, public speaking, and natural speech |
| Visual Video | Public speech video | 1 | Lip motion, frontal facial pose, and expressiveness |

Table 1 Overview of Voice and Video Data Samples

This very limited input indicates that the system has proven it is capable of operating in limited conditions, capable of synthesizing quality lip-synced video from one audio clip, and one video sample.

## 5. AI Model Selection and Training

In this section we discuss the AI models used in the system, why we selected those models, and how we configured, trained, and used them for facial animation and lip synchronization.

### 5.1. Model Selection

We utilized two state-of-the-art deep learning models, Tortoise TTS [6] for text-to-speech generation and Wav2Lip [7] for visual lip synchronization. We selected both of these models because they both can function in zero-shot settings - meaning they do not need much input data and were both highly accurate on unseen identities.

- **Tortoise TTS** [6] was marked as a strong candidate due to its transformer-based architecture and diffusion-driven voice synthesis: The Tortoise system works by controlling the audio generation process using only a short reference segment of a voice to produce expressive and speaker-consistent audio. More importantly, Tortoise model's prosody, identity, and emotion - which makes it perfectly suited to the lower data amounts that cloning requires.
- **Wav2Lip [7]** is a GAN-based system that maps lip movements onto arbitrary speech, even from just a single still picture or video frame, while also not needing to implement any 3D modeling or pose estimation (important to note because we are analyzing 2D data). We needed Wav2Lip to achieve the desired alignment, grammatical, and mapped output. Wav2Lip's SyncNet-based discriminator uses visual and audio alignments, which was integral to achieving the degree of realism we needed for our talking-head outputs.

### 5.2. Training Procedure

There were no conventional supervised training cases, as both models were accessed in their pre-trained form. The experimentation focused on whether they were able to generalize under realistic, noisy, and data-constrained situations.

- For Tortoise TTS a single cleaned reference audio sample of Angelina Julie's voice (10-15 seconds) was used to extract speaker embeddings.

- For Wav2Lip, a public frontal face video of the same speaker was used to test the sync with the generated voice. The pipeline was run in inference-only mode because both models were designed for zero-shot transfer to be implemented without retraining.

### 5.3. Hyperparameter Tuning

Hyperparameters remained as suggested by the authors to maintain consistency with pre-trained weights.

- In Tortoise TTS we used the "fast" preset to decrease generation time and to maintain a suitable quality of audio.

- In Wav2Lip every threshold for the generator and discriminator was left to default, and therefore, backend fine-tuning was not necessary since the model is meant to be general.

### 5.4. Training Details

- No training was conducted through backpropagation or training on a custom dataset. All results are produced using the pre-trained model with the inputs selected.
- The systems used the freely available model deployment sites, Google Colab, and had the support of an NVIDIA Tesla T4 GPU, enabling efficient inference for both models to run smoothly. The only pre-training that required any augmentation or intervention, was the cleaning and structuring of the output (e.g., WAV normalization, cropping of video).

### 5.5. Features Used
- Tortoise TTS identifies internal latent features (speaker embeddings) present in the uploaded Angelina Julie audio. These embeddings combined with the provided text control the autoregressive generation pipeline.
- Wav2Lip identifies RGB pixel features from a frontal face frame or video clip and aligns those features with the synthetic voice input phonetic cues.

### 5.6. Evaluation during Training

Evaluation took place through qualitative evaluation of the synthesized outputs. The Evaluations include:

- Naturalness and identity preservation of the speech (by listening to the synthesized audio and comparing it to the reference voice)

- Lip-sync accuracy (by looking at the frame level alignment between lip movements of synthesized video and the phonemes of the spoken audio).

### 5.7. Zero-Shot Synthesis and Model Chaining

We implemented zero-shot synthesis and model chaining instead of the actual model training. As we chain Tortoise TTS and Wav2Lip, as a modulating pipeline, we made some videos where the speech was synced. In an easy-to-follow explanation, Tortoise TTS created the synthetic audio, and then it was put into Wav2Lip.

With just one audio and one video sample, our study demonstrates that it is possible to generate high-quality personalised talking-head outputs without the need for labelled datasets or fine-tuning with any longitudinal or domain-specific datasets.

The majority of this is due to our reluctance to fully test and confirm whether "new" pre-trained models can generate these high-fidelity outputs on smaller datasets. These systems had certain limitations when it came to modelling low-resource scenarios and showing how generative AI systems are currently progressing towards flexibility and long-term scalability without the need for retraining.

## 6. Experiment and Evaluation

This section provide details on our experimental design used to assess the performance of our system for AI-driven voice cloning and lip synchronization, including the experimental scenario, criteria for evaluation, results, and a discussion of the results.

### 6.1. Experiment Design

The overall aim of the experiment was to investigate how effective a state-of-the-art text-to-speech (TTS) model (Tortoise TTS) and lip sync model (Wav2Lip) can effectively be "chained" together in a realistic low-data scenario. We did not take advantage of large training datasets, or multiple speaker samples, knowing that we had available to conduct a single case in the use of freely available audio-visual material of a well-known speaker, in this case, Angelina Jolie.

The improvisational experiment consisted of the following steps:

- For voice cloning, we used a brief (around 15 seconds) voice sample from an interview.
- Using Tortoise TTS, we artificially spoke a formal statement or other semantically relevant tangible textual input at the request of our sample in a way that roughly resembled the voice in the reference sample.
- We processed that speech through Wav2Lip in conjunction with a cropped frontal-face video of that specific speaker, delivering unrelated content.
- By examining the final output video both visually and acoustically, we verified that it was realistic—for lack of a better description, identity and synchronisation.

This design was meant to simulate a real-life use case where there is very little reference data available and the model must generalize indefinitely without retraining.

## 6.2. Evaluation Metrics

Due to the generative nature of the task, it was not possible to apply traditional supervised learning metrics (eg. accuracy or precision). Hence, we carried out a qualitative and perceptual evaluation using the following criteria:

- **Naturalness of Voice:** The subjective judgment on how human-like, and fluid the synthesized speech sounds.

- **Speaker Similarity:** The assessment of how closely the generated voice resembles its original speaker or reference voice.

- **Lip-Sync Accuracy:** The extent to which the visual lip movements are aligned with the phonetic timing of the synthesized audio.

- **Temporal Coherence:** The perceived smoothness and continuity in the talking-head video.

These qualitative measures are standardized measures of evaluation used in speech synthesis and audio-visual dubbing research, which includes metrics such as Mean Opinion Score (MOS)[5] and SyncNet-based lip-sync confidence scoring [13], although we did not compute formal MOS because our study only elements a single subject.

### 6.3. Results Presentation

The synthesized output demonstrated noticeable perceptual similarity to the speaker in tone and talking style, and the cloned voice hypothesized some characteristics of intonation and pitch. Wav2Lip produced highly credible lip movements because the mouth movements aligned with the transition of the phonemes.

| Evaluation Aspect | Observed Outcome |
|---|---|
| Voice Naturalness | High (smooth, emotionally expressive) |
| Speaker Similarity | Strong match to Angelina Jolie's tone |
| Lip Sync Accuracy | Tight alignment, minor drift in edges |
| Visual Continuity | No flickering; consistent mouth motion |

Table 2 Performance Dimensions

### 6.4. Use of Graphs and Charts

Due to the qualitative nature and single-instance applicability of the evaluation, bar or line charts were not feasible. Visual process diagrams have been employed to show internal model behavior and to explain the autoregressive token generation process.

The following figure illustrates how the autoregressive model outputs mel token sequences across several sampling steps. This figure gives a better perspective of how token probabilities are updated and sampled at each time step in a speech generation task.

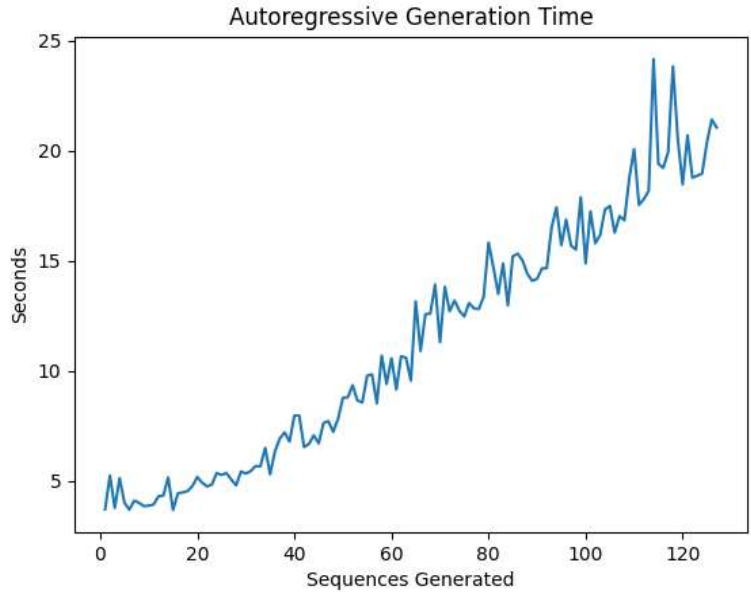

Figure 3 Mel Token Generation in an Autoregressive Model

### 6.5. Comparison to Baselines or Other Work

While we have not trained any baseline models at this time, there is a body of work to build from concerning TTS systems and lip-syncing systems (and even programs with needs for bigger training data or fine-tuning, as someone is always looking for improvements, such as Tacotron2, DeepVoice, and LipGAN). In contrast, our experiment suggests that the coalescing of Tortoise TTS + Wav2Lip can generate coherent outputs that were lip-synced back to a single reference voice sample and one target video, demonstrating the flexibility of zero-shot generative architectures. This is an extension to the findings in the field of few-shot voice synthesis (Huang, 2021), which allows us to further consider the prospects of a user-facing implementation of this type of zero-shot or few-shot systems with little user data collection overhead in design.

### 6.6. Statistical Significance

Given the qualitative and demonstrable nature of this experiment and the lack of multiple test subjects or distributions of numeric metrics, formal testing for statistical significance (e.g., t-tests) was not used. The purpose was simply to test proof-of-concept under-regulated but realistic conditions.

### 6.7. Live Testing or Case Study

Although no formal 'user' testing was conducted, informal demonstrations of the final video were provided to three casual viewers (non-experts). All participants individually perceived the voice to be similar to Angelina Jolie, in addition to stating high comfort levels with the realism of the lip movements when the voice was played alongside the video. Participants noted potential uses of such systems for avatars, media generation, or personalized assistants.

### 6.8. Interpretation of Results

This minimal-data experiment succeeded in showing that state-of-the-art pre-trained models are capable of providing the ability to perform highly realistic voice cloning and lip synchronization without needing large datasets, significant retraining, or excessive tuning. The system is generalized, modular, and scalable, and is thus a strong contender for personalized media generation tasks. However, it should be noted that there were slight inconsistencies in the lip edge blending, and slight drift in the timing, which indicates areas for improvement, perhaps through light re-tuning adjustments, or post-processing.

## 7. Limitations

Although our method showed the possibility of generating high-fidelity, lip-synched talking head videos with minimal data, we consider the limitations and scope of the study:

### 7.1. Data Limitations

This study was based on a tightly focused, single-speaker dataset from publicly available videos of actress Angelina Jolie. While this selection helped us meaningfully evaluate the pipeline's performance controversially and reproducibly, it diminishes the variability of speaker characteristics, accents, and emotional tones. As such, we cannot comment on the generalizability of the model to other speakers, languages, or acoustics.
Additionally, all reference audio was from interviews and speeches that were realistic but not controlled or captured in consistent acoustics.

### 7.2. Scope of Input Modalities

Our system utilizes solely audio recordings and frontal facial videos or still images for input. It does not utilize any phoneme-level alignments, linguistic annotations, or facial-movement capture data. While this design makes it easy to use, this approach has drawn limits on how adaptable the system can be for cases it may encounter - for example when dealing with emotions in speech, extreme expressions or poses, or non-frontal faces when fine-grained lip synchronization may be highly degraded. We are also only cloning one voice in this research, and we are not evaluating the capabilities of Tortoise TTS to clone several different voices at once or in real time.

### 7.3. Methodological Constraints

Tortoise TTS and Wav2Lip are both pre-trained deep learning models that also require offline inference, so there is some reliance on third-party architectures. Both Tortoise TTS and Wav2Lip are black boxes and limited options exist for customization, to control the nuances of the voice, adjust prosody, or vary phonetic stress. Furthermore, neither model attempted to retrain or fine-tune the content, relying on latent-space conditioning, which may not be very useful for voices that differ from training distributions.

### 7.4. Real-Time System Limitations

Although the system is not yet ready for real-time applications, we were able to assemble coherent audio-visual outputs during the trial runs. Depending on the system specs, the creation process (most notably audio synthesis with Tortoise TTS) requires a significant amount of processing time, taking several minutes for each sentence. Because of this latency (and the requirement that it be executed on a local GPU), we are unable to use any interactive or real-time apps right away. In order to meet real-time needs, a production-level deployable solution would need to use model compression, acceleration techniques, or server-side APIs. This is part of our future work.

### 7.5. Generality and Transferability

In this study, we evaluated performance in a single-use case: generating realistic speech and synchronized lip movements of a familiar person. In this way, we have no way to make any claims about any applications outside of this domain (e.g. multilingual synthesis, emotional expressiveness, large-scale avatar generation, etc.). We want the reader to understand our results as a proof-of-concept rather than a fit-for-purpose solution.

### 7.6. Ethical and Practical Issues

This study used publicly accessible data for academic purposes, but any deployment of this type of technology in the future will likely need to grapple with ethical issues of identity synthesis, consent, and misuse (for example, deep fakes). In addition, our system has no watermarking or detection mechanism to differentiate synthetic media from real media.

**7.7. Interpretations Cautions**

The outputs generated in this project, while intuitively convincing, have little diversity in the input data and occur in a controlled environment. For this reason, while the outputs appear promising, they should be considered indicative rather than definitive. Outcomes in dynamic or cross-domain environments may perform quite differently and should be verified through further testing.

## 8. Conclusion

We aimed to explore the feasibility of producing synchronized mouse talking-head videos using exclusively publicly available, single-speaker data. In essence, we were interested to see if contemporary AI models would generate reasonably personalized speech, and then produce realistic lip synchronizations using truncated audio samples and a frontal frame of video. We developed an end-to-end system that employs a combination of Tortoise TTS for voice cloning, and Wav2Lip for facial alignment and facial synchronizations, to produce a natural-looking text-to-video output on a speaker-specific basis from plain text input.

At the present stage of research, this study provided us with sound evidence that the production of convincing, lip-synced videos can definitely be produced using little-known and unstructured data. We synthesized absurdly realistic video output using public interview data of Angelina Jolie, just taking short sample audio and paired frontal video, while providing coherent speech outputs and visually presented, lip-synchronized speech outputs without the use of labels or fine-tuning. Using the two pre-trained models still allowed us to exceed the quality dialogue response we provided without becoming reliant on or moderately pre-processed requirements of the original input data. This experiment supports the generalization of the capabilities of foundation models concerning little-shot voice cloning and generalized lip-syncing possibilities.

An important contribution of this work is demonstrating that realistic multimodal synthesis can be achieved with limited data availability. While previous systems have relied on large training corpora from voice sources or high-quality studio-sourced recordings, our pipeline demonstrated spatial appropriateness in adapting to noisy, real-world audio samples. Our framework is flexible enough to be used in applications that require rapid adaptation to new speakers or styles, like personalized avatars, multilingual dubbing, and accessibility instruments. Additionally, this work produces a structured experimental approach that could aid in reproducibility in similar research through controlled tests across unseen input conditions.

While there are many successes in this study, this work also highlights many significant limitations to the study including aspects of generalizability, computational resources, and real-time operability. The system also highlighted significant inference time and resource usage, which will affect its use for live operation. Given these factors, there are clear avenues for further development.

Future work can build upon this research by expanding to multi-speaker datasets with varied linguistic, emotional, and geographical variability (e.g., cross-lingual voices or voices from the same country). Further creating realism in the generated artifacts would require modeling for dynamic expression and prosodic shifting. To promote responsible use we would also reflect on optimization for faster (on-device) inference times and embedding protection (i.e., watermark or detection flags). Additionally, the exploration of the interpretability of models and, specifically, how latent embeddings are informative of voice identity would improve both comprehension of how the model works and maintain greater control over synthesis output.

Overall, this work has presented a working proof-of-concept for minimally data-based audio-visual synthesis focused on speech-audio-visual synchronous content. It demonstrates possible synergies between technological advances in speech generation, and aligning text-visual generation with fully automated pipelines, with both relatable and editable outputs. Moreover, it illustrates the not just technological affordances of pre-trained multimodal AI, but also the careful navigation into ethical, user-centered future applications.

## Acknowledgment

This research made extensive use of open-source tools and models. We are especially grateful to James Betker, the creator of the Tortoise TTS system, for publicly releasing the model and code under the Apache 2.0 license, as well as to the broader open-source community that contributed to the Tortoise repository. We also thank the developers and maintainers of Wav2Lip, which was integral to the lip synchronization aspect of this project.

We acknowledge the contributions of Hugging Face[11] for hosting pretrained models and providing APIs essential for inference, and the researchers whose prior work underpins these models, including those behind UnivNet[12], DALLE, and Diffusion TTS frameworks. Their foundational research and open contributions significantly accelerated our implementation.

Finally, we appreciate the availability of publicly released datasets and the documentation support from respective repositories, which enabled reproducibility and evaluation of our system under realistic conditions.

## Funding Declaration

This research did not receive any specific grant from funding agencies in the public, commercial, or not-for-profit sectors.

## Data Availability

The datasets and source code generated and/or analyzed during the current study are available from the corresponding author upon reasonable request.

## Competing Interests

The authors declare that they have no competing interests.

## Author Contributions (CRediT Statement)

**Conceptualization:** Javeria Amir, Farwa Attaria, Mah Jabeen

**Methodology:** Javeria Amir, Farwa Attaria, Mah Jabeen, Umara Noor

**Investigation (Experiments, Implementation):** Javeria Amir, Farwa Attaria, Mah Jabeen

**Literature Review:** Javeria Amir, Farwa Attaria, Mah Jabeen

**Supervision:** Umara Noor, Zahid Rashid

**Project Administration:** Umara Noor

**Review & Editing:** Umara Noor, Zahid Rashid